\documentclass[reprint,amsmath,amssymb,aps]{revtex4-1}
\usepackage{graphicx}
\usepackage{dcolumn}
\usepackage{bm}
\usepackage{wrapfig}
\usepackage{soul,xcolor}
\usepackage{physics} 
\usepackage{mathtools}
\usepackage{array}
\usepackage{comment}
\usepackage{makecell,float}
\begin{document}

\title{First-principles characterization of the magnetic properties of Cu$_2$(OH)$_3$Br}

\author{Dominique M. Gautreau}
\affiliation{Department of Chemical Engineering and Materials Science, University of Minnesota}
\affiliation{School of Physics and Astronomy, University of Minnesota}
\author{ Amartyajyoti Saha}
\affiliation{Department of Chemical Engineering and Materials Science, University of Minnesota}
\affiliation{School of Physics and Astronomy, University of Minnesota}
\author{Turan Birol}
\affiliation{Department of Chemical Engineering and Materials Science, University of Minnesota}

\date{\today}

\begin{abstract}
Low dimensional spin-1/2 transition metal oxides and oxyhalides continue to be at the forefront of research investigating nonclassical phases such as quantum spin liquids. 
In this study, we examine the magnetic properties of the oxyhalide $\text{Cu}_2\text{(OH)}_3\text{Br}$ in the botallackite structure using first-principles density functional theory, linear spin-wave theory, and exact diagonalization calculations. This quasi-2D system consists of $\text{Cu}^{2+}$ $\mathrm{S} = 1/2$ moments arranged on a distorted triangular lattice.
Our exact diagonalization calculations, which rely on a first-principles-based magnetic model, generate spectral functions consistent with inelastic neutron scattering (INS) data. By performing computational experiments to disentangle the chemical and steric effects of the halide ions, we find that the dominant effect of the halogen ions is steric in the $\text{Cu}_2\text{(OH)}_3\text{X}$ series of compounds. 
\end{abstract}

\maketitle


\section{\label{sec:intro}Introduction}

Low-dimensional magnetism can give rise to a host of exotic phenomena. For example, in quasi-1D systems, the fractionalization of magnons into spinons has been both theoretically predicted and experimentally observed \cite{Coldea2001,Kohno2007,Mourigal2013}. In addition to low-dimensionality, frustration also plays a key role in the magnetic behavior of a system, as in the Majumdar-Ghosh model, which predicts a valence-bond solid \cite{Majumdar1969one,Majumdar1969two} or in the Kitaev honeycomb model, which predicts a spin liquid \cite{Kitaev2006}. 
Emergent phenomena in low-dimensional, and especially frustrated, magnets are promising for a wide range of applications including high-T$_C$ superconductors, quantum computation, and skyrmion memories \cite{Han2012, Leonov2015}. However, the ideal realization of theoretical models is hard to come by due to many reasons, including coupling to the lattice \cite{Savary2017, Chamorro2020, Paul2020}. This makes it essential to study low-dimensional magnetism in novel compounds.

In recent years, Kagome antiferromagnets have garnered extensive interest, especially the Herbertsmithite compound, which contains local spin-1/2 magnetic moments \cite{Norman2016}. Despite large antiferromagnetic interactions ($J \sim  17 $ meV) which couple nearest-neighbor spins, no signature of magnetic ordering is measured down to $\sim$~50~mK \cite{Helton2007}. This makes Herbertsmithite a prime candidate for hosting a quantum spin liquid ground state. Herbertsmithite can be formed from paratacamite Cu$_2$(OH)$_3$Cl by replacing one quarter of the Cu atoms with Zn atoms \cite{Inosov2018}. In this study, we shed light on the magnetic properties of a related series of compounds, the botallackite Cu$_2$(OH)$_3$X series, where $\mathrm{X}=\mathrm{Cl}$, Br, or I, with a particular emphasis on Cu$_2$(OH)$_3$Br.
 
Previous experimental studies \cite{Zheng2004,Zheng2005,Zheng2009} show that the spin-$1/2$ copper moments of $\text{Cu}_2\text{(OH)}_3\text{Br}$ order antiferromagnetically on a distorted triangular lattice. More recent experiments \cite{Zhang2020} have used inelastic neutron scattering (INS) to probe the excitation spectrum, and revealed a diffuse continuum above the magnon bands, which may indicate a multi-spinon or multi-magnon continuum. While the distorted crystal structure hosts parallel chains with ferromagnetic and antiferromagnetic spin ordering, a complete picture of the magnetic ground state and excited states of this compound remains unknown. A detailed first-principles theoretical study on the magnetic properties of Cu$_2$(OH)$_3$Br, and the origin of the observed magnetic excitation spectrum is to this day missing, with the exception of \cite{Zhang2020} where we used the first-principles-determined parameters to obtain a spin wave spectrum which agrees with the experimentally observed magnon bands. 

While all oxyhalide botallackite $\text{Cu}_2\text{(OH)}_3\text{X}$ compounds are antiferromagnets \cite{Zheng2009}, the N\'eel temperature changes dramatically from 7.2 K to 10 K to 14 K upon changing X from Cl to Br to I. This seemingly suggests that the dominant superexchange pathways involve the halogen ions \cite{Zheng2009}. However, this is at the same time counter-intuitive, as the Cu -- O distance is significantly smaller than the Cu--halogen distance \cite{Zheng2005}. This point has not been clarified by using first-principles methods yet. 

In this study, we elucidate the magnetic interactions as well as the resulting ground state and excitations in $\text{Cu}_2\text{(OH)}_3\text{Br}$ and related compounds. For this, we use density functional theory (DFT) to extract parameters of a minimal magnetic Heisenberg model, and we determine the excitation spectrum using linear spin wave theory and exact diagonalization. We repeat the first-principles calculations for different halide ions and different crystal structures to determine the effects of the crystal structure and the chemical differences between the halide ions on the magnetic interactions. Our results show that 1) even though the excitation spectrum is seemingly 1-dimensional, the magnetic Hamiltonian of $\text{Cu}_2\text{(OH)}_3\text{Br}$ is 2-dimensional, 2) the frustration between Cu ions on different 1-dimensional spin chains gives rise to the 1-dimensional-like behavior of the excitation spectrum, and 3) the predominant effect of the halide ion on magnetism is steric, i.e. different sizes of the halide ions change the crystal's structural parameters, leading to an indirect effect on the magnetic exchange interactions, and hence the N\'eel temperature. 

This paper is organized as follows: In the following section, we discuss the methods used for our calculations. In Sect. \ref{sec:structure}, we introduce the botallackite structure, the magnetic energy expression, and the exchange constants of our model. In the first subsection of Sect. \ref{sec:results}, we extract exchange coupling constants for the two-dimensional Heisenberg Hamiltonian and examine the classical magnetic ground state. In Subsect. \ref{sec:exchange_pathway}, we discuss whether magnetic interactions are predominantly mediated through superexchange via the halogens by performing calculations with different structures and ions. In Subsect. \ref{sec:holsteinmath}, we use the DFT-obtained exchange constants to calculate the magnon dispersion relation within linear spin wave theory (LSWT). In Sect. \ref{sec:ED}, we perform exact diagonalization calculations to probe the nature of the quantum excitations. We conclude with a summary. 


\section{\label{sec:methods}Methods}
 DFT calculations were performed using the Projector Augmented Wave approach as implemented in the Vienna Ab Initio Simulation Package (VASP)\cite{Kresse1993,Kresse1996CMS,Kresse1996PRB}. Results were obtained on a 2x1x1 supercell using a 4x8x8 k-point grid, along with the PBEsol approximation to the exchange correlation functional \cite{Perdew2008}. To properly reproduce the local magnetic moments on the Cu ions, the rotationally invariant LSDA+U scheme introduced by Liechtenstein et al. with $\mathrm{U}=\mathrm{4~eV}$ and $\mathrm{J}=\mathrm{0~eV}$ was used \cite{Liechtenstein1995}. The trends we report are qualitatively stable against variations of the U and J values within reasonable ranges. The reported results are obtained from collinear magnetic calculations. We also performed noncollinear calculations with spin-orbit coupling and obtained qualitatively very similar results, in line with the quenched orbital magnetic moments of the Cu ion. 

In Sect. \ref{sec:exchange_pathway}, for all comparisons between the different botallackite materials, we used the experimental structures with hydrogen atoms selectively relaxed, as the positions of the hydrogen atoms were not determined experimentally. This is in contrast to the results in Tab. \ref{tab:exchangeBr} for Cu$_2$(OH)$_3$Br, in which the experimental H positions are reported in the literature, and hence no atoms were relaxed.

We performed exact diagonalization calculations using our in-house python code, following the standard formalism of \cite{Sandvik2010}. To obtain eigenvalues and eigenvectors of our Hamiltonian matrices, we used the Lanczos algorithm as implemented in SciPy \cite{Lanczos1950,Virtanen2020,VanderWalt2011}.


\section{\label{sec:structure} Crystal Structure and Magnetic Model}

\begin{figure}
\centering
\includegraphics[width=0.8\linewidth]{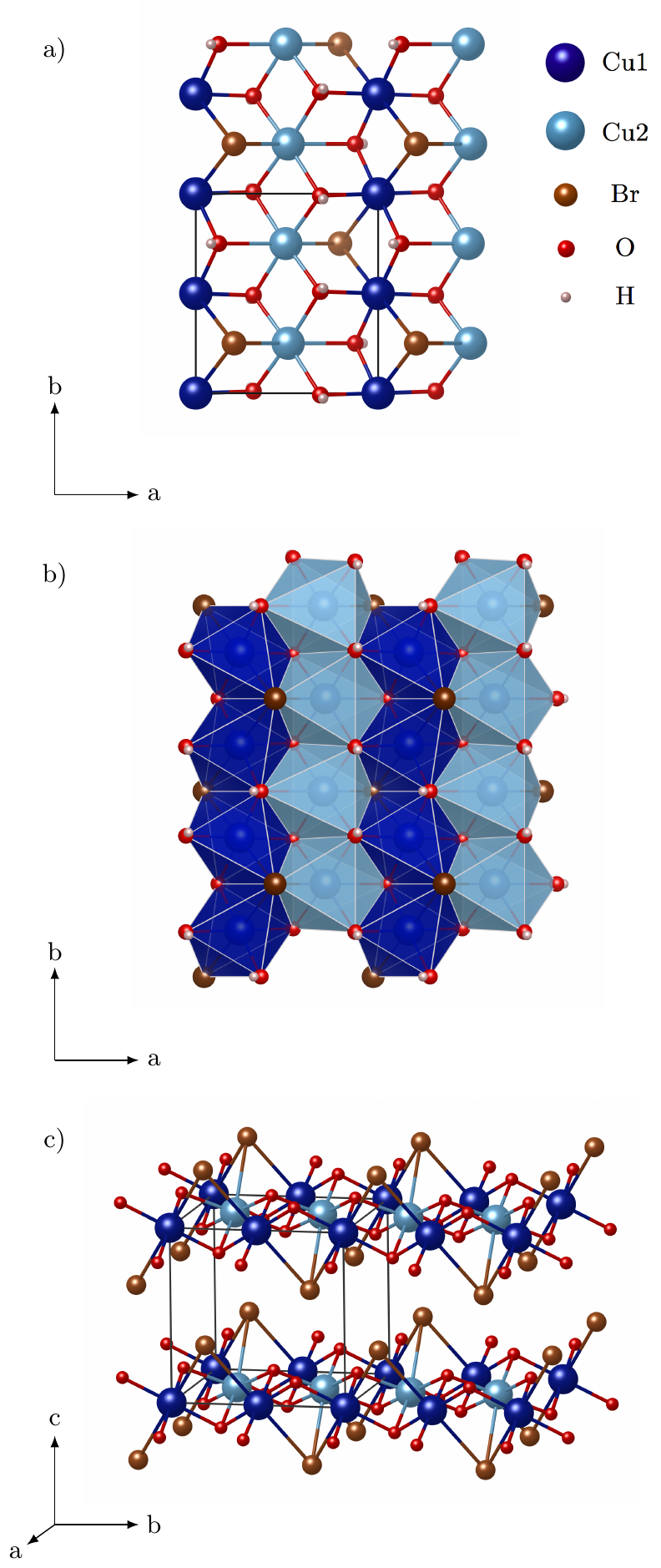}
\caption{a) A unit cell of the botallackite structure and (b) the same diagram but showing octahedra. c) The botallackite structure forms ab-planes.}
\label{fig:structure}
\end{figure}

The botallackite structure consists of weakly bound layers as shown in Fig. \ref{fig:structure}. It is monoclinic with the space group P2$_1$/m. Each Cu ion is octahedrally coordinated with anions and has electronic configuration $3d^9$, and therefore a net spin-$1/2$ moment. The two different types of Cu ions (with either one or two halide ions in their coordination octahedra) form a distorted triangular lattice. Each anion octahedron shares six of its edges with co-planar neighboring octahedra  (Fig. \ref{fig:structure}b). The planes are weakly bound to each other by hydrogen bonds (Fig. \ref{fig:structure}c). As a result, the electronic structure is very 2-dimensional, and we ignore all magnetic interactions between neighboring layers. 

\begin{table}
\begin{tabular}{c  c  c}
        \hline \hline
        Atom & Wyck. Pos. & Site Sym.\\
        \hline
        Cu$_1$  &2a     & $\bar{1}$ \\
        Cu$_2$  &2e     & $m$\\
        H       &2e     & $m$\\
        H       &4f     & $1$\\
        O       &2e     & $m$\\
        O       &4f     & $1$\\
        X(Br, Cl, I)       &2e     & $m$\\
        \hline \hline
\end{tabular}
        \caption{Wyckoff positions and site symmetries of the atoms in the Cu$_2$(OH)$_3$X botallackite with space group P2$_1$/m.}
\label{tab:wyckoff}
\end{table}

The Wyckoff positions for the atoms in space group P2$_1$/m are presented in Table \ref{tab:wyckoff}. There are two symmetry-inequivalent Cu atoms at Wyckoff positions 2a and 2e, which we denote as Cu$_1$ and Cu$_2$ respectively. The coordination octahedra surrounding the Cu$_1$ atoms are composed of four oxygen ions and two bromine ions. In contrast, the octahedra surrounding the Cu$_2$ atoms are composed of five oxygen ions and one bromine ion. 
The Cu$_1$ -- Cu$_1$ nearest neighbor superexchange pathways are formed from shared O -- Br edges, while the Cu$_2$ -- Cu$_2$ nearest neighbor superexchange pathways are formed by shared O -- O edges (Fig \ref{fig:exchangeAndgdstate}a). (Note that direct cation--cation interactions are often considerable in edge-sharing geometries as well \cite{Goodenough1963}.) We denote the exchange constants arising from the above pathways $J_1$ and $J_2$, respectively. Both Cu$_1$ -- Cu$_1$ and Cu$_2$ -- Cu$_2$ neighbors align along the $b$ crystallographic direction. 
The only nearest-neighbor exchange interactions along the $a$ direction are between Cu$_1$ -- Cu$_2$ pairs. There are two symmetry-inequivalent Cu$_1$ -- Cu$_2$ pairs, where the two octahedra share an edge containing either two oxygens, or an oxygen and a halogen ion. This leads to a significant difference between the exchange pathways and constants. We denote the exchange constants specifying these two inequivalent pathways $J_3$ and $J_4$ respectively. Thus, there are four nonequivalent nearest-neighbor Heisenberg couplings, as shown in Fig. \ref{fig:exchangeAndgdstate}a. We also consider exchanges $J_5$ and $J_6$, describing in-plane next-nearest neighbor Cu$_2$ -- Cu$_2$ and Cu$_1$ -- Cu$_1$ interactions, but find them to be negligible as discussed below. We do not consider any higher order exchange interactions, such as biquadratic or ring-like terms, since they are not expected to be strong in spin-1/2 systems \cite{Fazekas1999, Paul2020}. 
In other words, the magnetic model we consider is a Heisenberg model on a distorted triangular lattice:
 \begin{equation}
	\mathcal{H}= E_0+\sum_{<ij>}J_{ij} \textbf{S}_i \cdot \textbf{S}_j.
	\label{equ:heis}
\end{equation}
We determine the paramagnetic energy $E_0$ and the exchange constants $J_{ij}$ by linear regression from DFT energies. We consider a larger number (19) of magnetic configurations compared to the number of independent parameters in Eq. \ref{equ:heis} to ensure that there is no over-fitting despite the number of free parameters in the model. The final sets of parameters reproduce the DFT energies with small errors and so no other terms are needed in the magnetic model. Additional information regarding the calculation of the spin exchanges, as well as the relative importance of the next-nearest-neighbor exchanges can be found in the Supplementary material \cite{Supplement}.

\section{\label{sec:results}Results}


\subsection{\label{sec:MagGdState}Classical magnetic ground state}

The exchange values obtained from the DFT+U calculations for the model, including exchanges with nearest and next-nearest neighbors, are listed in Table \ref{tab:exchangeBr}.
The next-nearest neighbor exchange constants $J_5$ and $J_6$ are both as small as 0.1~meV, which is comparable to the statistical error of the fit. (The systematic error of the DFT calculations is likely larger.) Furthermore, the values of $J_1$ through $J_4$ are not affected within the statistical error bars when we exclude $J_5$ and $J_6$ from our model. As a result, we henceforth consider only the nearest-neighbor exchanges $J_1$ through $J_4$.

\begin{table}
\centering
\begin{tabular}{|c|c|c|c|}
\hline
\hline
 & Exchange 		& Shared 	& Cu--Cu distance \\
 & constant (meV) 	& anions 	& (\AA)\\
\hline
$J1$ & -1.4 $\pm$ 0.1& 1 Br, 1 O &  3.074  \\
$J2$ & 4.8 $\pm$ 0.1  & 2 O &  3.075 \\
$J3$ & 0.8 $\pm$ 0.1  & 2 O &  3.174  \\
$J4$  & 0.4 $\pm$ 0.1 & 1 Br, 1 O&  3.237  \\
$J5$  & 0.1 $\pm$ 0.1 & -- &  5.624 \\
$J6$  & 0.0 $\pm$ 0.1 & -- &  5.624 \\
\hline
\hline
\end{tabular}
\caption{Exchange constants for Cu$_2$(OH)$_3$Br in the experimentally determined structure}
\label{tab:exchangeBr}
\end{table}

\begin{figure}
\centering
\includegraphics[width=0.90\linewidth]{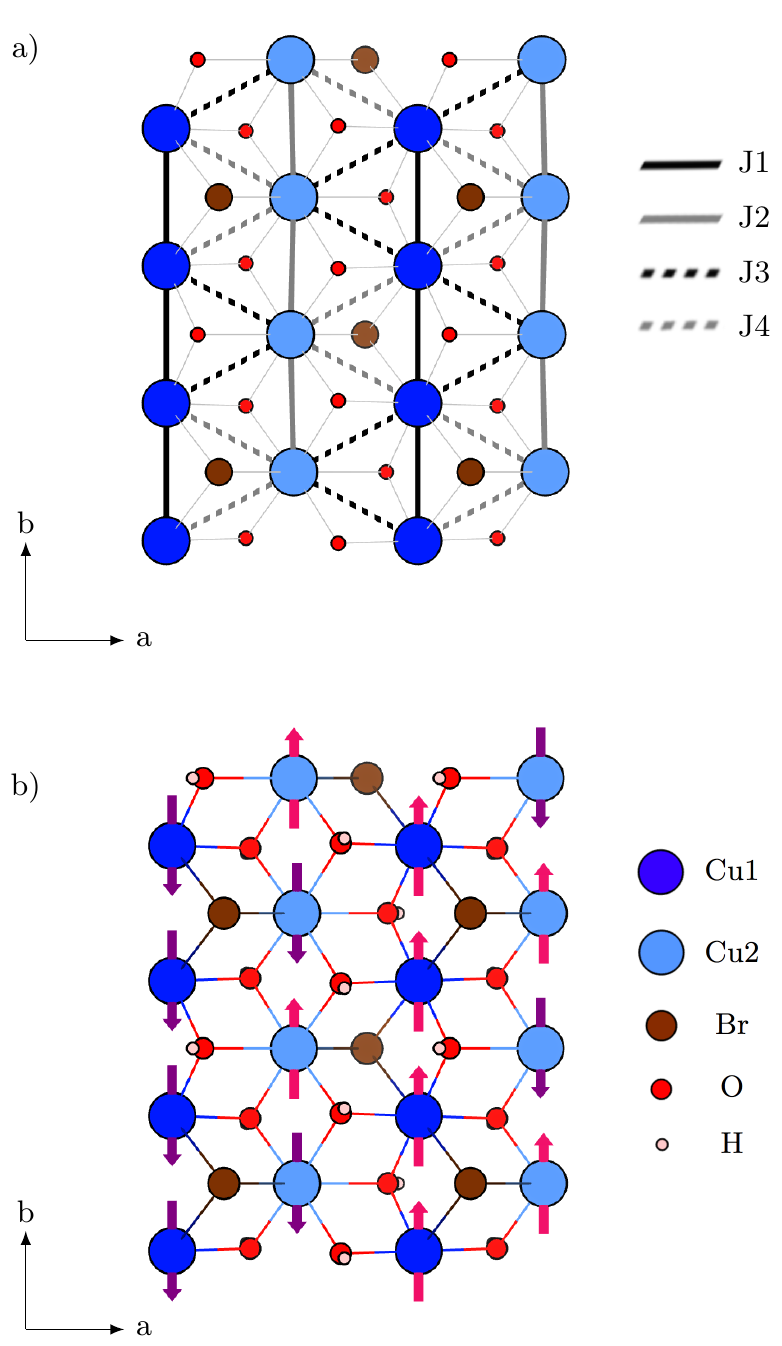}
\caption{a) Definitions of the magnetic exchange constants and b) the proposed classical magnetic ground state.}
\label{fig:exchangeAndgdstate}
\end{figure}

Regarding the nearest-neighbor couplings, we see that $J_1$ and $J_2$ are the largest magnitude exchange constants and are negative and positive, respectively. This gives rise to alternating ferromagnetic and antiferromagnetic chains which extend along the $b$-direction, in line with the experimental observations. Although the interchain couplings $J_3$ and $J_4$ are small, they are still relevant, in contrast to the negligible $J_5$ and $J_6$ couplings. This seems to contradict the results of inelastic neutron scattering results \cite{Zhang2020} which at first glance suggest that $J_3$ and $J_4$ are zero, since the spin-wave dispersion is flat in the interchain direction. However, as we discuss below, the inclusion of $J_3$ and $J_4$ preserves the relatively flat bands in the interchain direction. 

The classical magnetic ground state can be determined by allowing the exchange constants to be satisfied in order of descending magnitude. Neutron measurements show collinearity within chains and canting between spins on different chains \cite{Zhang2020}. To include the effect of this noncollinearity, we would have to include anisotropic spin exchange and single ion anisotropy terms in the Hamiltonian which arise due to spin-orbit coupling. These terms are difficult to extract from DFT, due to the low symmetry of the botallackite compounds and the weakness of SOC therein. Our noncollinear DFT calculations that include spin-orbit coupling show that the effect of magnetocrystalline anisotropy is negligible compared to the symmetric nearest-neighbor exchanges. We ignore the effect of spin-orbit coupling in the rest of this study, and assume a collinear magnetic order.

We begin with satisfying $J_1$ and $J_2$, which results in ferromagnetic and antiferromagnetic chains extending along the $b$-direction, as discussed above. Next, the interchain coupling $J_3$ constrains the relative orientation of spins on neighboring chains, leading to the classical magnetic ground state shown in Fig. \ref{fig:exchangeAndgdstate}b. In this configuration, the spins  on every other chain alternate in direction. This magnetic configuration agrees qualitatively with neutron scattering experiments \cite{Zhang2020}. We emphasize that the nonzero interchain coupling $J_3$ is necessary to reproduce the experimentally observed long-range magnetic order. We separately determine via the Luttinger-Tisza method \cite{Luttinger1946,Litvin1974} that this result is the exact classical magnetic ground state of our model \cite{Supplement}.


\subsection{\label{sec:exchange_pathway}Superexchange interactions}

The N\'eel temperatures of the botallackite cuprates change substantially (from 7.2 K to 10 K to 14 K) when substituting different halogen atoms X into $\text{Cu}_2\text{(OH)}_3\text{X}$ (from X = Cl to Br to I) \cite{Zheng2009}. This observation was interpreted as evidence of the halogen ions providing the dominant exchange pathways between the copper magnetic moments. However, the crystal structure itself suggests otherwise, since the copper--oxygen distances are much smaller than the copper-halogen distances. For example, in $\text{Cu}_2\text{(OH)}_3\text{X}$, while the copper--oxygen distances vary from around 1.9 to 2.3 \AA, the copper--halogen distances are much larger at 2.88 \AA. In this section, we present the first-principles projected density of states and the spin density in real space in order to examine the influence of the crystal structure and halogen ion type on the magnetic exchange parameters. Our results suggest that the dominant contribution to exchange is from direct hopping between the Cu ions and superexchange through the oxygen ions. 
Furthermore, we find that the largest effect of the different halogen ions is steric. That is, the halogen ions modify the magnetic interactions through changes in the crystal's structural parameters.

\begin{figure}
    \centering
    \includegraphics[width=0.9\linewidth]{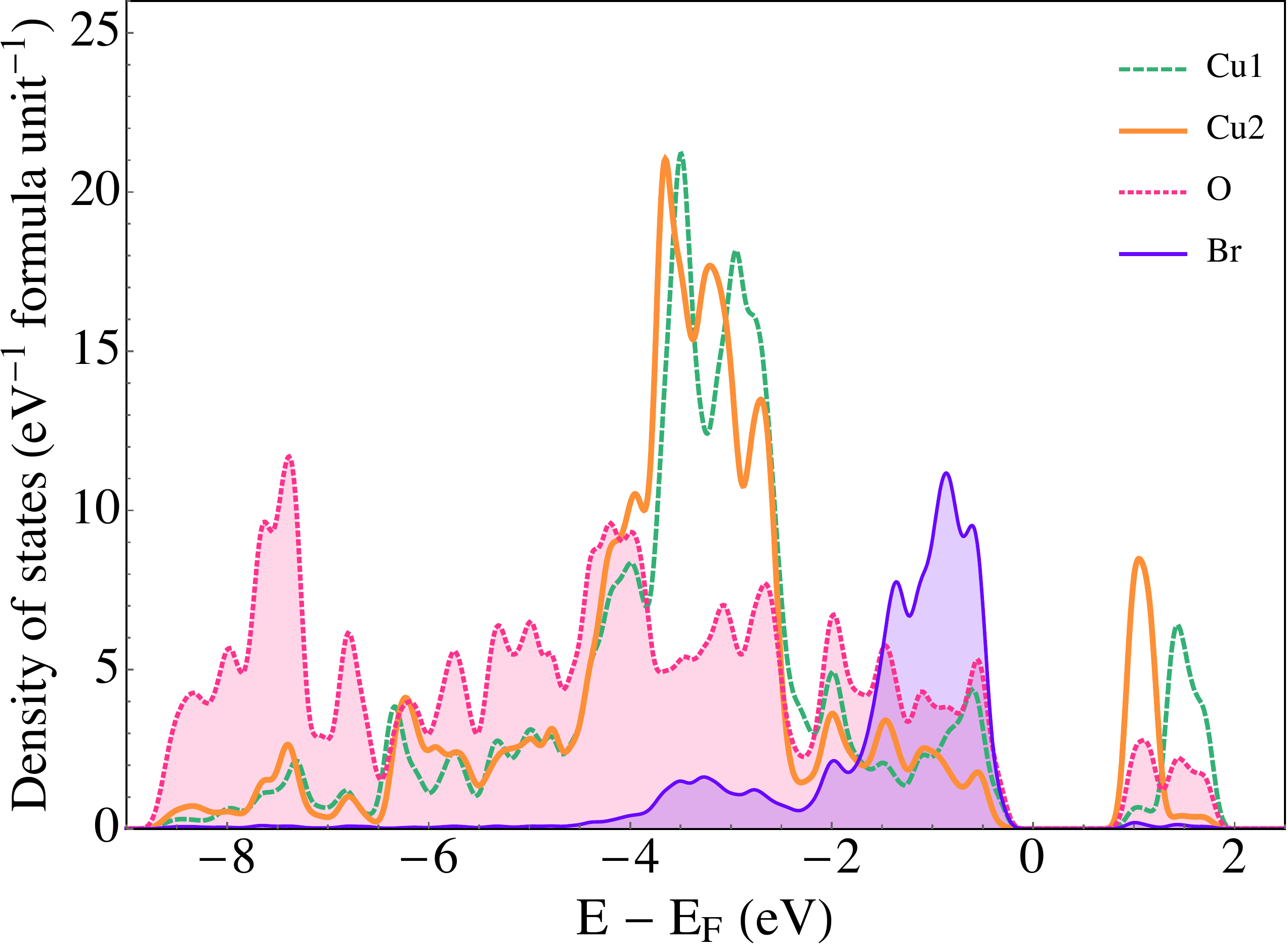}
    \caption{The projected density of states for the system in the antiferromagnetic ground state. The low energy DOS above the Fermi level has only a small contribution from the halogen states.}
    \label{fig:pdos}
\end{figure}

Fig. \ref{fig:pdos} shows the projected density of states (DOS), obtained from DFT calculations, for the bromide system in the proposed classical magnetic ground state. 
The Cu$^{2+}$ ions each have a single hole in the d-shell, which corresponds to peaks in the DOS between 1-2~eV above the Fermi level. For the halogen-mediated superexchange to be significant, the Cu -- Br hoppings would need to be large, which would lead to hybridization (mixing) of the spin-polarized Cu orbitals with Br--p orbitals. 
This Cu -- Br hybridization would appear in the DOS as Br weight in the same energy range as the low-lying, unoccupied Cu DOS. However, there is seemingly no Br weight in the DOS in this energy range, despite the high DOS of Br ions at the top of the valence band. This implies that superexchange is predominantly oxygen-mediated, as expected from the geometrical and crystal field considerations explained below.

\begin{figure}
\centering
\includegraphics[width=0.49\textwidth]{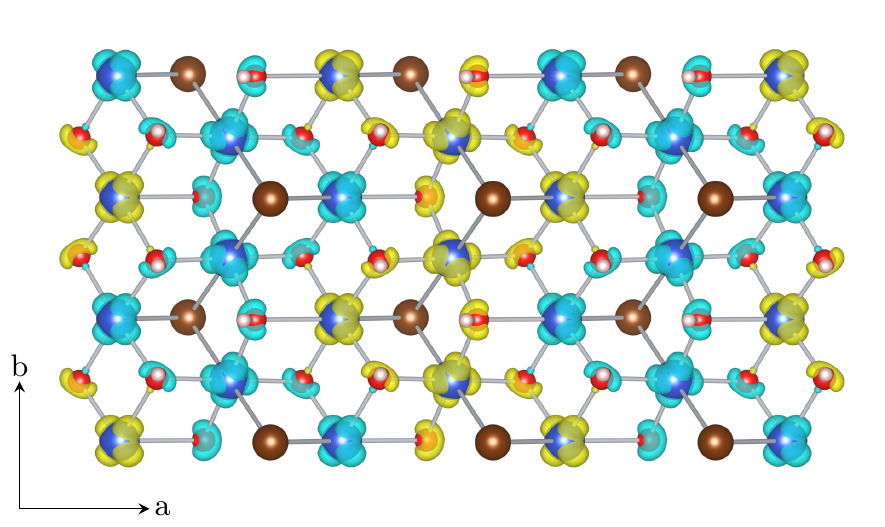}
\caption{Isosurface of spin density, which shows orbital ordering, in the antiferromagnetic ground state. Positive and negative spin density are denoted by blue and yellow, respectively.}
\label{fig:spindensity}
\end{figure}

In order to clarify the nature of hybridization between the Cu and O states, we explore the system's `orbital ordering'. 
While the Cu ions are octahedrally coordinated, the site-symmetries are low enough to split the $e_g$ orbitals. The different charges of the ligands enhance the effects of these low-symmetry crystal fields. As a result, while it is not possible to talk of a proper orbital ordering or a transition between orbitally ordered and disordered states, the alignment of the partially occupied d-orbital on each Cu atom can shed light on the source of the magnetic interactions. 
Fig. \ref{fig:spindensity} shows the isosurface of spin density obtained from a DFT calculation in the antiferromagnetic state. When using local coordinate axes on each Cu ion, with the $z$-axis pointing towards the Br ions, the spin-polarized, half-filled orbitals have $d_{x^2-y^2}$ character and lobes pointing towards the oxygen atoms. This orbital occupation preference can be understood simply by electrostatic contributions to the crystal field: Br has -1 charge, as opposed to -2 of O, and the Cu-Br bond lengths ($\approx$ 3\AA) are significantly larger than the the Cu-O bond lengths ($\approx$ 2\AA). As a result, the electrons in the $d_{2z^2-x^2-y^2}$ orbital experience a lower electrostatic repulsion from the Br anions, and thus have lower energy \footnote{In passing, we note that the double occupation of the $d_{2z^2-x^2-y^2}$ orbital would result in a greater electrostatic repulsion on the Br ions, which would in part be the cause of the longer Cu--Br distances as well. This can be considered in a similar vein as the Jahn--Teller effect, though in the botallackites the symmetry is already broken by the inequivalent anions.}.

\begin{figure}
\centering
\includegraphics[width=0.95\linewidth]{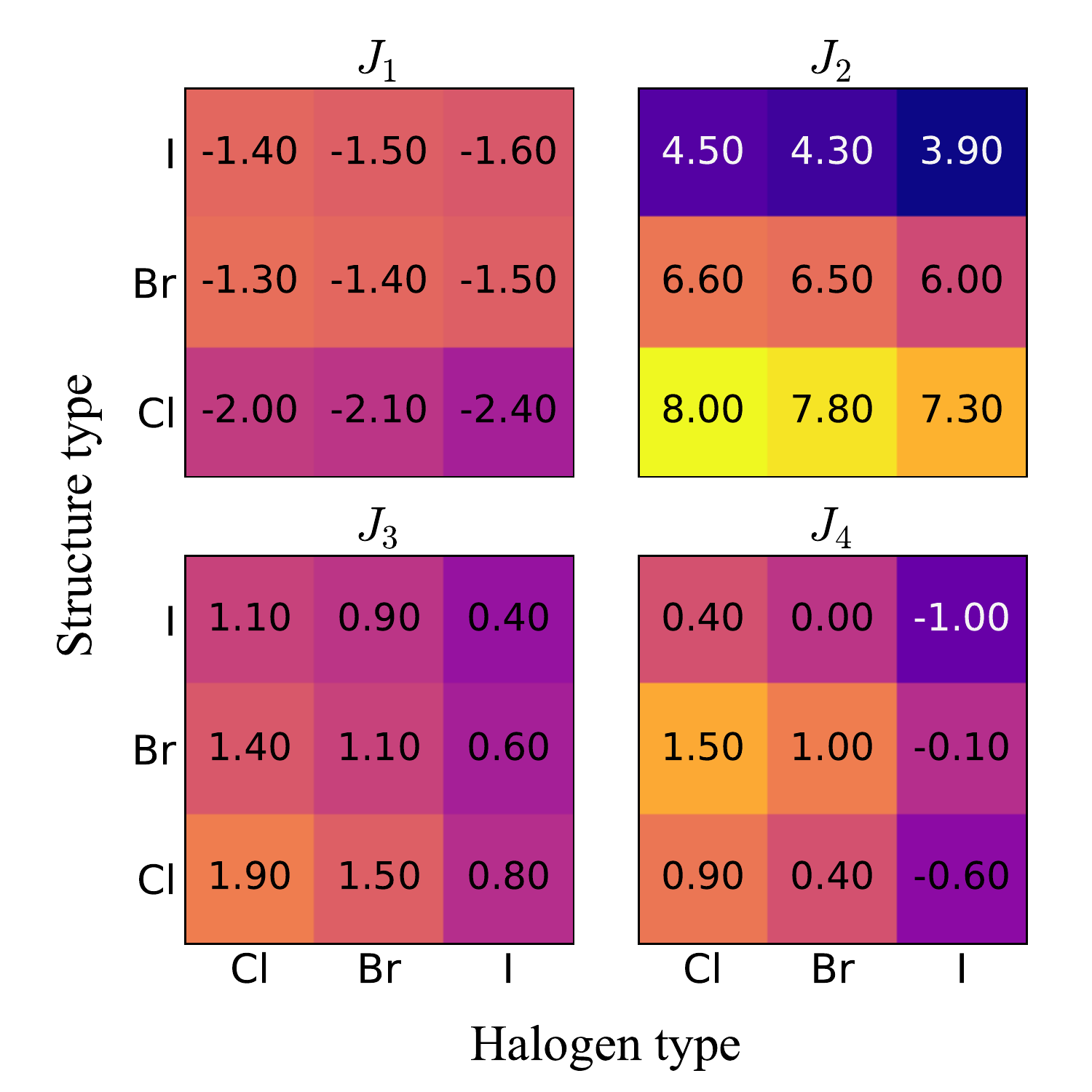}
     \caption{Color maps which show how the exchange constants vary with insertion of different halogen types into a given structure \cite{Note2}. Within a row of a given color map, the structure type is fixed. Moving from column to column corresponds to substituting different halogen atoms into this fixed structure.}
    \label{fig:color_map_exchanges}
\end{figure}

The improper orbital ordering imposed by the strongly asymmetric crystal field leads to superexchange interactions dominated by Cu--O--Cu pathways: the oxygen p-orbitals sigma bond with the half-filled Cu $d_{x^2-y^2}$ orbitals, which leads to a large hopping element $t_{O-Cu}$. In contrast, the bromine p-orbitals, which point to the center of the Cu $d_{x^2-y^2}$ orbitals, have very small hopping $t_{Br-Cu}$ due to the equal but opposite contributions to the hopping integral from different lobes of the $d_{x^2-y^2}$ orbital. (Note that we are only considering the hopping elements $t=\langle \psi_{Cu}|\mathcal{H}|\psi_O\rangle$ of the $d_{x^2-y^2}$ orbital, since the virtual excitations from the other, fully occupied orbitals to oxygen are forbidden by the Pauli principle.) Therefore, the Br ions likely have little role in mediating the magnetic exchange. This is substantiated by the lack of any visible spin-density on the Br ions, unlike the O ions, in Fig. \ref{fig:spindensity}.
 
Our findings and arguments so far suggest that the halogen cations do not directly mediate magnetic superexchange, which appears to contradict the large differences between the different botallackites' N\'eel temperatures observed experimentally \cite{Zheng2009}. However, an attribute of the different halogens that we have not yet considered is their different ionic sizes, which lead to changes in the lattice parameters as well as Cu--O--Cu angles. To disentangle the direct and indirect (crystal structure related) effect of the halogen ions on magnetism, we perform the following first-principles `computational experiments': 1) Using the experimentally obtained $\mathrm{Cu_2(OH)_3Br}$ structure, which we have thus far used to obtain the exchange constants $J_1$ through $J_4$, we repeat the DFT calculations to obtain the exchange constants, but with chlorine or iodine ions replacing the bromine ions. These calculations display the direct effect of the halogens without taking into account the indirect, steric effects of different ionic sizes. 2) We then use the experimentally measured  $\mathrm{Cu_2(OH)_3Cl}$ and  $\mathrm{Cu_2(OH)_3I}$ crystal structure parameters, and calculate $J_1$ through $J_4$ for each ionic compound again. Comparing the $J$ values of a particular compound in different crystal structures shows the strength of the indirect effect, i.e. the effect of the crystal structural changes on magnetism. 
The results of these calculations are shown in Fig. \ref{fig:color_map_exchanges}. Note that the $\mathrm{Cu_2(OH)_3Br}$ structure used in these comparison calculations is not identical to the experimental structure used previously \footnote{In the Cl and I compounds, the experimental data did not include the positions of the hydrogen atoms. Therefore, we have selectively relaxed the hydrogen atoms to obtain the lowest-energy structures. For consistency in the comparison plots, we then also relaxed the hydrogen atoms in the Br compound. We would like to point out that this relaxed-hydrogen structure is thus slightly different from the Br structure obtained experimentally, leading to different exchange couplings, as calculated by DFT.}. 

The results in Fig. \ref{fig:color_map_exchanges} indicate that the intrachain couplings $J_1$ and $J_2$ are relatively unaffected upon substituting different halogen ions while keeping the crystal structure fixed. The largest difference in $J_1$ and $J_2$ for a fixed structure is no more than 20\%. On the other hand, changing the crystal structure while keeping the stoichiometry fixed has a much stronger effect on these intrachain exchanges, which can be seen as a steeper color gradient in the vertical direction. In light of the arguments we introduce above, this result is not surprising. Since the predominant effect of the halogen ion is steric, we expect halogen-mediated exchange to be insignificant. The effect of halogen type is especially weak for $J_2$, which is the interaction between two Cu ions that share only O ligands.
As a result, we conclude that the intrachain couplings overall are not halogen mediated, but rather depend on a combination of direct exchange and superexchange through the oxygen sites \footnote{We can gain insight into the nature of the intrachain interactions by further examining this improper orbital order. From Fig. \ref{fig:spindensity}, we see that the Cu$_1$ chains have antiferro-orbital order while the Cu$_2$ chains are ferro-orbitally ordered. From the Goodenough-Kanamori-Anderson rules, we then expect ferromagnetic coupling between Cu$_1$ spins within a chain and antiferromagnetic coupling between Cu$_2$ spins within a chain. This indeed agrees with the exchange couplings we extract from DFT, in which we find $J_1 < 0$ and $J_2 > 0$.}.

The trends are less clear and opposite for the interchain exchanges $J_3$ and $J_4$. For both $J_3$ and $J_4$, the changes are smaller in absolute terms compared to the intrachain exchanges, but larger in relative terms. Also, changing the halogens leads to a stronger difference than changing the crystal structure. $J_3$ decreases by more than a factor of two upon substitution of Cl with I. This is especially surprising, because the $J_3$ interaction couples two Cu ions that do not share a halogen ligand.  
The only sign change is observed in $J_4$, and only for compounds containing I. The different sign of $J_4$ in the iodide compound does not lead to a different classical magnetic groundstate, but relieves the frustration between the interchain exchanges.
This stronger dependence of the interchain exchanges on the halogen ions can be explained by the fact that the direct exchange, which exponentially depends on the Cu-Cu distance, contributes less to $J_3$ and $J_4$. The interchain nearest-neighbor Cu ions are farther from each other (by around 0.1 to 0.2~\AA), and thus any change in the anion-mediated superexchange is more important.

Although we have shown that the largest exchanges in the system, $J_1$ and $J_2$, are not halogen-mediated, we note that the exact way in which the N\'eel temperature depends on the exchange constants is unclear. In the extreme limit of 1D compounds that host chains with no shared ligands, such as the A$_3$BB'O$_6$ family \cite{Bergerhoff1956, Nguyen1995, Birol2018}, two separate transitions with temperatures $T_1$ and $T_2$ may be observed, which are determined by intrachain and interchain interactions respectively. However, there is only a single transition observed in the botallackites. It is likely that the dominant exchange couplings $J_1$ and $J_2$ are not the main energy scales which determine the N\'eel temperature for this transition. Instead, $T_N$ is likely determined by the much weaker interchain couplings $J_3$ and $J_4$, along with the interplanar coupling(s). This would explain the two-fold difference between $T_N$ of the different oxyhalide botallackites.


\subsection{\label{sec:holsteinmath}{Linear spin wave theory}}

\begin{figure}
\centering
\includegraphics[width = 0.47\textwidth]{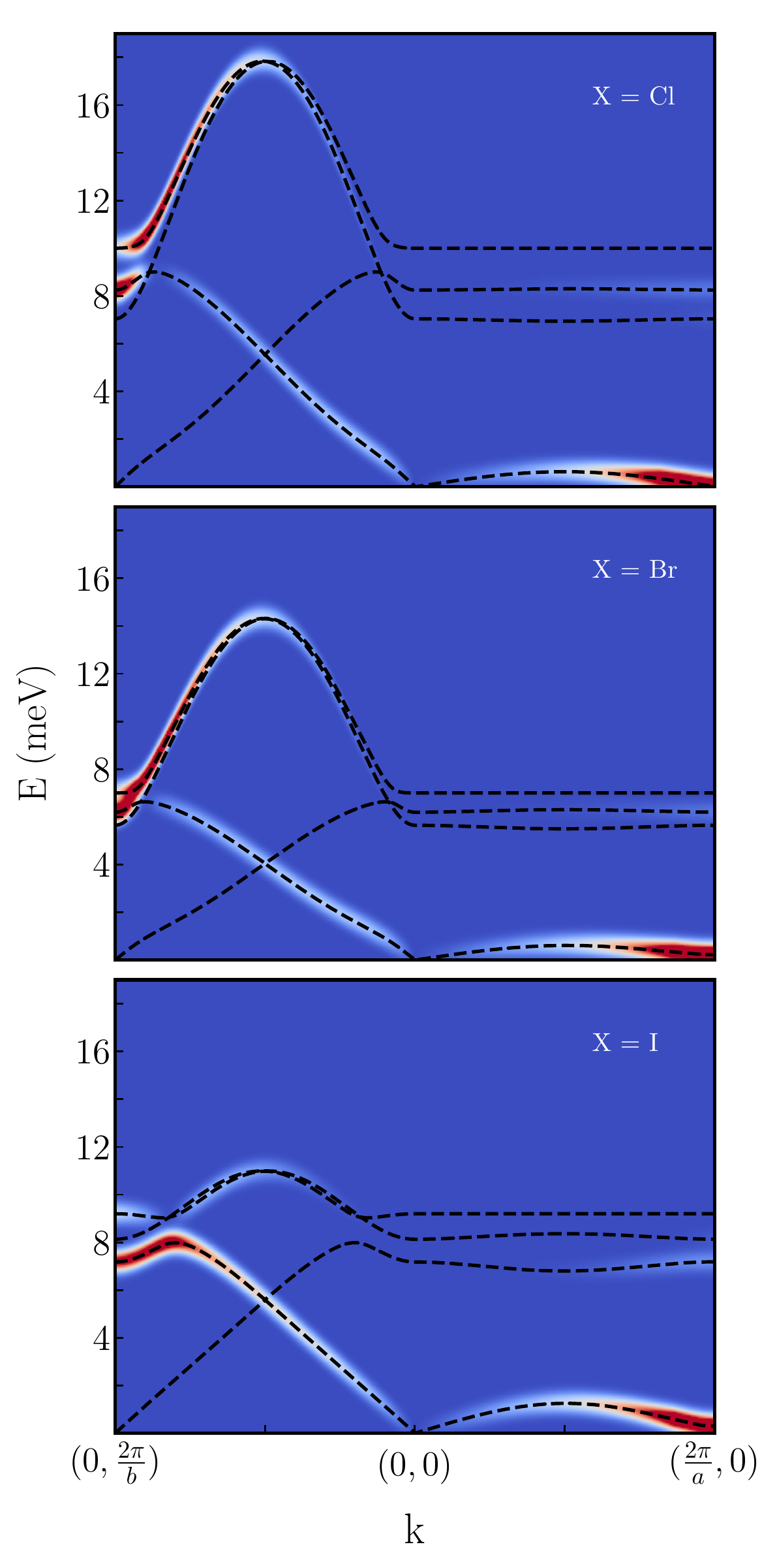}
     \caption{Results of the linear spin wave calculations. In addition to the dynamical structure factor $S(\mathbf{k},\omega)$, the dashed lines show the dispersions of the 4 magnon modes. At each k-point, only 2 of the 4 modes have non-negligible spectral weight. The structure factors are broadened by a Gaussian of width 0.3 meV. The exchange constants used are obtained for the structures with relaxed hydrogen atoms (see discussion at the end of Sect. \ref{sec:exchange_pathway}).}
     \label{fig:halogenbands}
\end{figure}

We now apply the Holstein-Primakoff transformation \cite{Fazekas1999} to our nearest-neighbor Heisenberg Hamiltonian to determine the spin wave spectrum of Cu$_2$(OH)$_3$X, for X=Cl, Br, and I. In this calculation, we take the classical magnetic ground state to be that which we determined from DFT calculations in Sect \ref{sec:MagGdState} (Fig. \ref{fig:exchangeAndgdstate}b), and the exchange parameters determined using the hydrogen-relaxed structures.

The results for our calculations are shown in Fig. \ref{fig:halogenbands} (dotted lines). We find that the bands along the intra-chain ($k_y$) direction are dispersive, while the bands along the inter-chain ($k_x$) direction are relatively flat, despite the nonzero values of $J_3$ and $J_4$.
The relatively flat dispersion in the $a$-direction can be partially explained as due to the smaller values of the interchain couplings $J_3$ and $J_4$ compared to the intrachain couplings $J_1$ and $J_2$. However, another contributing factor is the competition between $J_3$ and $J_4$ in the chloride and bromide compounds. Indeed, despite having a smaller average of $|J_3|$ and $|J_4|$, the iodide compound has the largest bandwidth in the interchain direction (a,0). This is due to the fact that $J_4$ is ferromagnetic in the iodide compound, in contrast to the chloride and bromide compound. There is therefore no competition between $J_3$ and $J_4$, and as a result no frustration in the iodide compounds. 
To further test this idea, we set $J_4=0$ in the chloride and bromide compounds, relieving the frustration in the system, and recalculate the magnon bandstructure. This leads to an increase in the bandwidths along $a$ axis as expected \cite{Supplement}.

The dynamical structure factor $S(\mathbf{k},\omega)$ is given by  
\begin{align}
    S(\mathbf{k},\omega) &=
    \sum_{\alpha}\int \frac{dt}{2\pi}e^{i\omega t}
    \langle \psi_0|S_{-\mathbf{k}}^\alpha(t)S_{\mathbf{k}}^\alpha(0)|\psi_0\rangle \\
    &=
    \sum_{n,\alpha} \abs{\langle \psi_{n,\mathbf{k}}|S_{\mathbf{k}}^\alpha|\psi_0 \rangle}^2 \delta(\omega -E_n(\mathbf{k})+E_0)
\end{align}
We calculate $S(\mathbf{k},\omega)$ within linear spin wave theory following Ref. \cite{Toth2015} and show the result in Fig. \ref{fig:halogenbands} for the three different $\mathrm{Cu_2(OH)_3 X}$ structures. The magnetic unit cell has 8 atoms, and as a result, there are four doubly degenerate magnon modes. However, at each wavevector, two of the four bands have negligible weight. This agrees with the neutron scattering data, which observes only two groups of excitations which are ferromagnetic and antiferromagnetic in character.


\subsection{\label{sec:ED}Exact diagonalization}

Although linear spin wave theory is a useful tool to gain insight into the magnon excitations of spin systems, there are other excitations which cannot be probed using this method. Multi-magnon states require higher-order corrections to take into account the magnon-magnon interactions. Additionally, there are other, lower spin excitations such as spinons, which become important in low-spin systems. For example, the 1D antiferromagnetic Heisenberg chain shows no magnons in its excitation spectrum, and hosts only spinon excitations, which are fractionalized spin-1/2 objects. 

Previous INS work claims to observe the coexistence of spinons and magnons, localized on the antiferromagnetic and ferromagnetic chains respectively in $\text{Cu}_2\text{(OH)}_3\text{Br}$ \cite{Zhang2020}. Recent INS work on another spin-1/2 cuprate, SeCuO$_3$ also points to a variety of coexisting excitations \cite{Testa2020}. In order to understand the INS results on  $\text{Cu}_2\text{(OH)}_3\text{Br}$ better, and shed light on the causes behind this apparent coexistence, it is necessary to perform calculations that can capture excitations beyond the scope of linear spin wave theory.

For this reason, we performed exact diagonalization calculations. Exact diagonalization involves constructing the exact Hamiltonian matrix for a finite size cluster with periodic boundary conditions and Heisenberg interactions, which we then diagonalize numerically \cite{Sandvik2010}. This, in principle, gives the exact spectrum of a material. In particular, multi-magnon and spinon states can be captured via exact diagonalization. 

\begin{center}
\begin{figure}
    \includegraphics[width=0.47\textwidth]{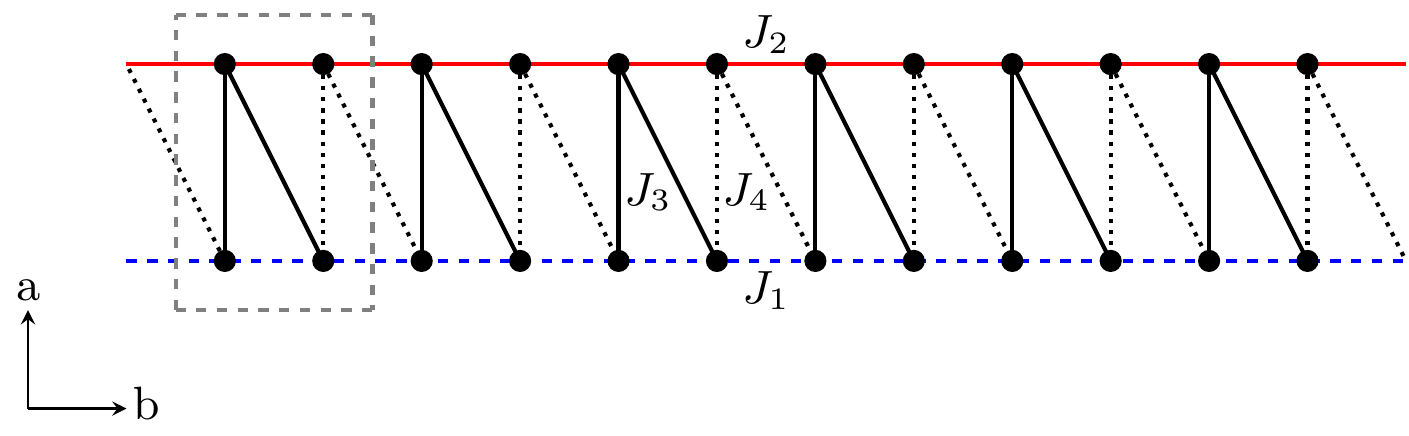}
\caption{The 24-site cluster used in our exact diagonalization calculations, extending six unit cells in the $b$-direction and one unit cell in the $a$-direction.}
\label{fig:cluster}
\end{figure}
\end{center}

For our exact diagonalization calculation, we use a 24-site cluster that extends six unit cells in the $b$-direction, and one unit cell in the $a$-direction, shown in Fig. \ref{fig:cluster}. The cluster is comprised of one ferromagnetic chain and one antiferromagnetic chain, each aligned along the $b$-direction. We impose periodic boundary conditions in the $b$-direction, and open boundary conditions in the $a$-direction. When obtaining the spectrum of our Hamiltonian, we exploit the SU(2) symmetry of our Heisenberg Hamiltonian and block diagonalize by $S_z^\mathrm{tot}$. Additionally, we use the translation symmetry in the $b$-direction to block diagonalize by crystal momentum $k$. After implementing these symmetries, the resulting Hamiltonian block containing the ground state is $\sim23,000\times23,000$. The Lanczos algorithm was used to find both the energies $E_n(\mathbf{k})$ and the corresponding weights $\abs{\langle \psi_{n,\mathbf{k}}|S_{\mathbf{k}}^\alpha|\psi_0 \rangle}^2$. 

\begin{figure}
\centering
    \includegraphics[width=0.48\textwidth]{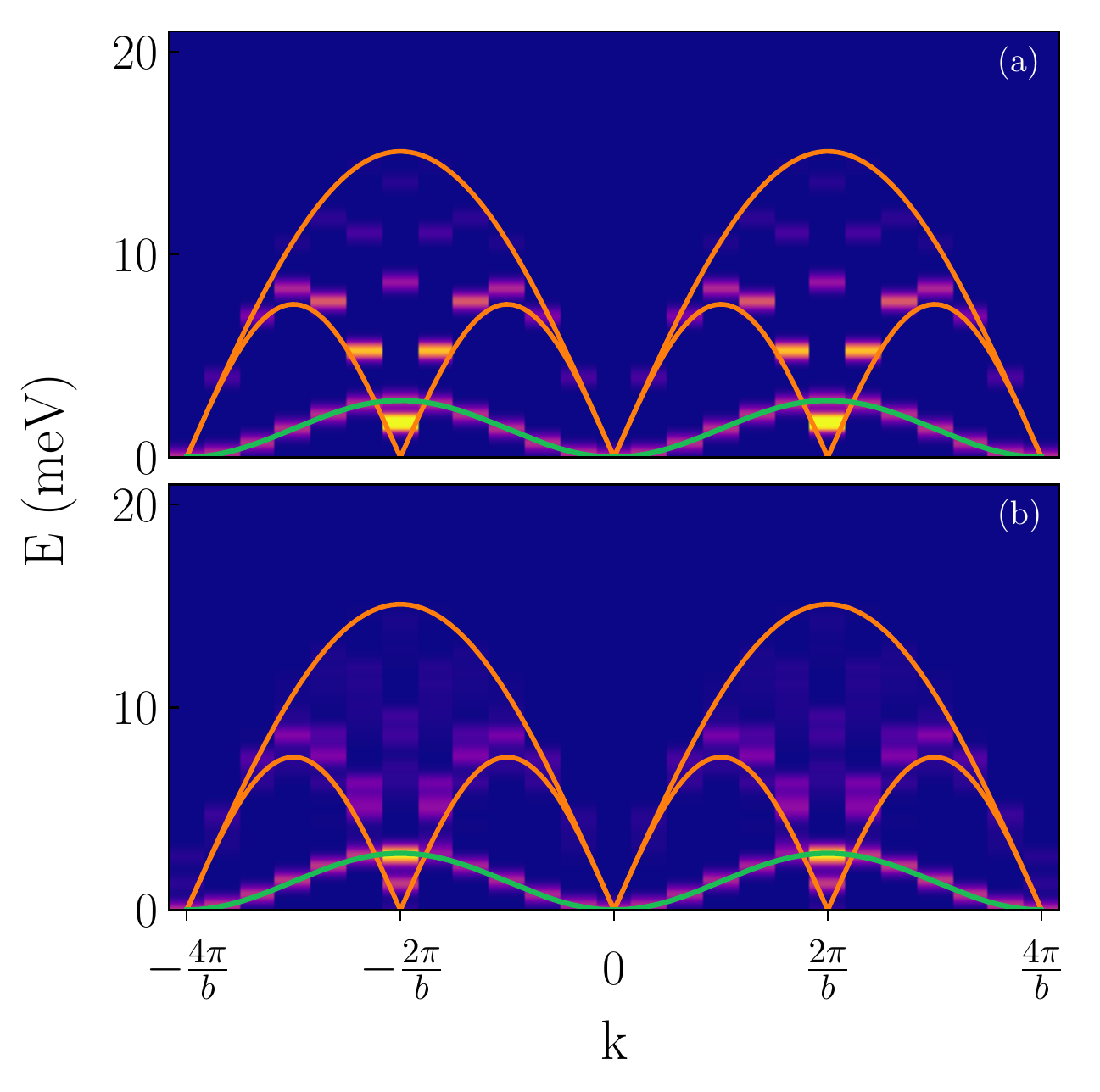}
    \caption{The dynamical structure factor $S(\mathbf{k},\omega)$ calculated using exact diagonalization for $\mathrm{Cu_2 (OH)_3Br}$ in the case where (a) there is zero interchain coupling and (b) the interchain couplings $J_3$ and $J_4$ are set to their values obtained from DFT. Also plotted are the bounds of the 2-spinon continuum (orange) obtained from the Bethe ansatz where $J_{AFM}$ = $J_2$ and the magnon band (green) for a 1D ferromagnetic Heisenberg chain where $J_{FM}$ = $J_1$. The exchanges were found for the $\mathrm{Cu_2 (OH)_3Br}$ structure obtained from experiment, without any relaxation of atoms.}
    \label{fig:dsf_ed}
\end{figure}

To investigate the effect of the interchain coupling, we calculate $S(\mathbf{k},\omega)$ for both decoupled chains ($J_3 = J_4 = 0$) and coupled chains, where $J_3$ and $J_4$ are the values extracted from DFT. The results are shown in Fig. \ref{fig:dsf_ed}, where we have used the exchange parameters obtained from the experimentally-determined Br compound (Tab. \ref{tab:exchangeBr}.) Here, in addition to plotting $S(\mathbf{k},\omega)$, we also plot the upper and lower bounds of the 2-spinon continuum as determined from the Bethe ansatz with $J_{AFM}$ = $J_2$ (orange lines) and the magnon band for a 1D ferromagnetic Heisenberg chain with $J_{FM}$ = $J_1$ (green line).

As expected, in the case of zero interchain coupling, we simply have the superposition of a magnon band from the ferromagnetic chain, and the two-spinon continuum from the antiferromagnetic chain (Fig. \ref{fig:dsf_ed}a). That is, spinons and magnons coexist in isolated chains of the material. Once the interchain couplings are introduced, the material no longer hosts isolated one-dimensional AFM chains, and thus should not have a pure 2-spinon continuum. Instead, we expect the magnon and spinon excitations to mix with each other. Indeed, as we increase interchain coupling, the spectral weight inside the 2-spinon continuum smears out significantly (Fig. \ref{fig:dsf_ed}b). 

\begin{figure}
\centering
    \includegraphics[width=0.45\textwidth]{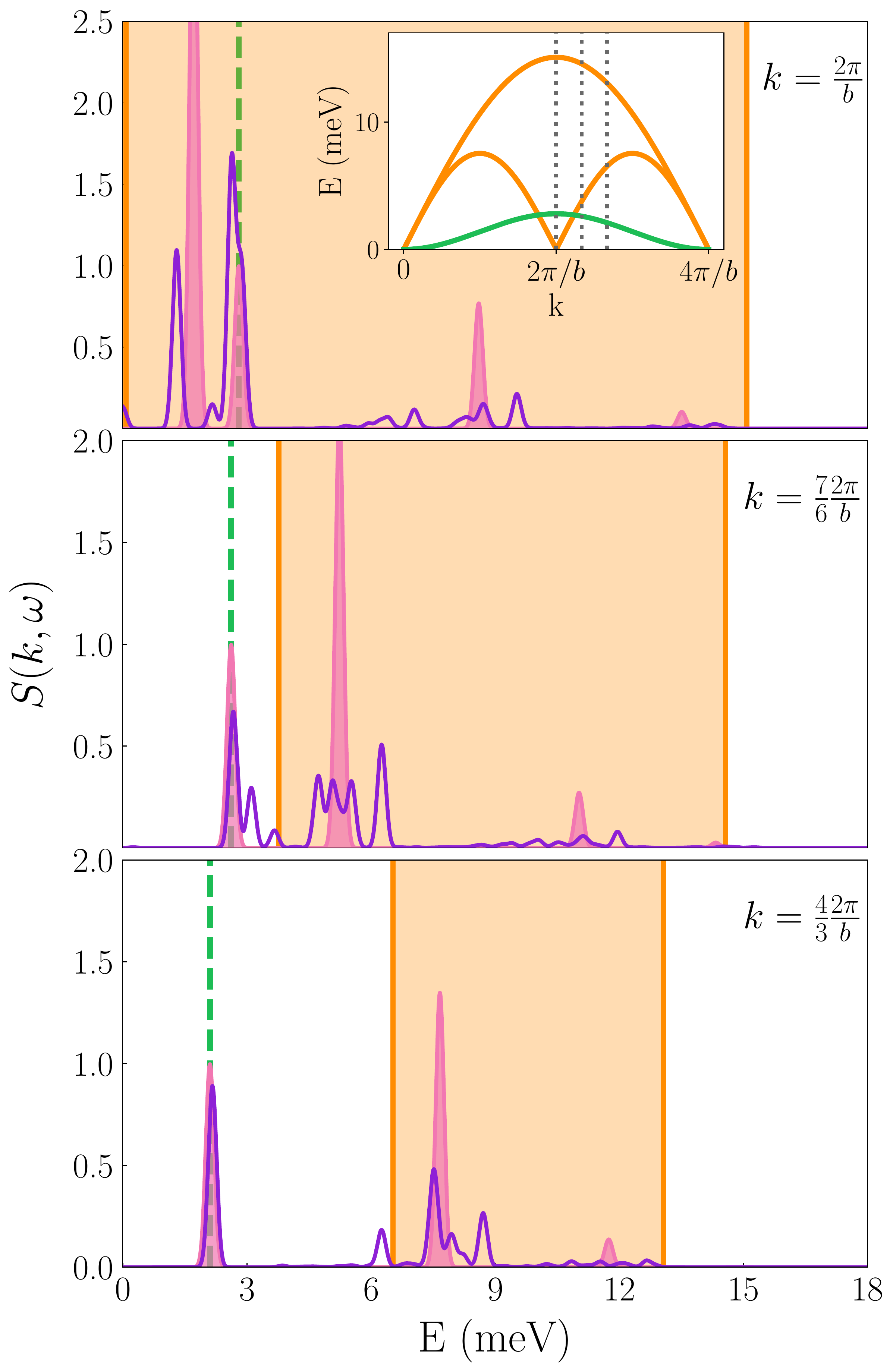}
    \caption{The dynamical structure factor $S(\mathbf{k},\omega)$ at three momenta: $k=\frac{2\pi}{b}$, $k=\frac{7}{6}\frac{2\pi}{b}$, and $k=\frac{4}{3}\frac{2\pi}{b}$, as shown in the inset of the top panel. In each panel, the dashed green line shows the position of the magnon band for a 1D ferromagnetic Heisenberg chain with $J_{FM}$ = $J_1$, while the shaded orange region delineates the bounds of the spinon continuum, as obtained from the Bethe ansatz, with $J_{AFM}$ = $J_2$. The results for the case of zero interchain coupling are shaded pink, while the results for the case of nonzero interchain coupling are shown in purple.}
    \label{fig:linecut}
\end{figure}

To highlight the effect of introducing interchain coupling, we plot $S(\vb{k},\omega)$ as a function of $\omega$ for both the coupled and decoupled chains (Fig. \ref{fig:linecut}) at several momenta ($k=\frac{2\pi}{b}$, $k=\frac{7}{6}\frac{2\pi}{b}$, and $k=\frac{4}{3}\frac{2\pi}{b}$), which are shown in the inset. In each panel, the position of the ferromagnetic band is shown as a dashed green line, and the bounds of the 2-spinon continuum are shaded in orange. The results for $S(\vb{k},\omega)$ for zero interchain coupling are shaded in pink, while $S(\vb{k},\omega)$ for nonzero interchain coupling is shown in purple. 

For $k = \frac{4}{3}\frac{2\pi}{b}$ (bottom panel), we find that the introduction of nonzero interchain coupling leads to a smearing of the spectral weight inside the 2-spinon continuum and a transfer of some weight outside of the bounds predicted by the Bethe ansatz \cite{Bethe1931,Karbach1998}. However, because the ferromagnetic band and the continuum are well-separated in energy, the ferromagnetic band is only weakly perturbed by the presence of nonzero interchain coupling. In the middle panel ($k=\frac{7}{6}\frac{2\pi}{b}$), the energy of the ferromagnetic band is closer to the 2-spinon continuum, and the effect of nonzero interchain coupling is more pronounced. Here, as with $k = \frac{4}{3}\frac{2\pi}{b}$, there is transfer of weight outside the bounds predicted by the Bethe ansatz. However, there is also appreciable mixing between the magnon and spinon excitations. In this energy range where mixing occurs, between 2 meV and 4 meV, we expect magnon-spinon interactions to become significant. Finally, for $k = \frac{2\pi}{b}$, we see that the ferromagnetic band overlaps with the spinon continuum. From the above arguments, we expect magnon-spinon interactions to become appreciable at this momentum. Indeed, we observe a substantial modification of the spectral weight at this intersection due to the presence of nonzero interchain coupling. However, what this implies for the character of the magnon and spinon excitations has yet to be determined.

Despite the presence of magnon-spinon interactions, the ferromagnetic magnon band is distinct from the spinon continuum for most wavevectors, which is consistent with the INS results \cite{Zhang2020}, because of the difference of the magnitudes of $J_1$ and $J_2$. As a result, although there is mixing between magnon and spinon excitations, qualitatively distinct magnon-like and spinon-like excitations coexist for most wavevectors in $\text{Cu}_2\text{(OH)}_3\text{Br}$. In other words, although the magnetic model for this material is two-dimensional, the system's magnetic excitations retain a one-dimensional character, because of a combination of the relative strengths of intrachain interaction, the competition between the interchain couplings $J_3$ and $J_4$, as well as the weakness of these couplings compared with $J_1$ and $J_2$.

\begin{figure}
\centering
    \includegraphics[width=0.45\textwidth]{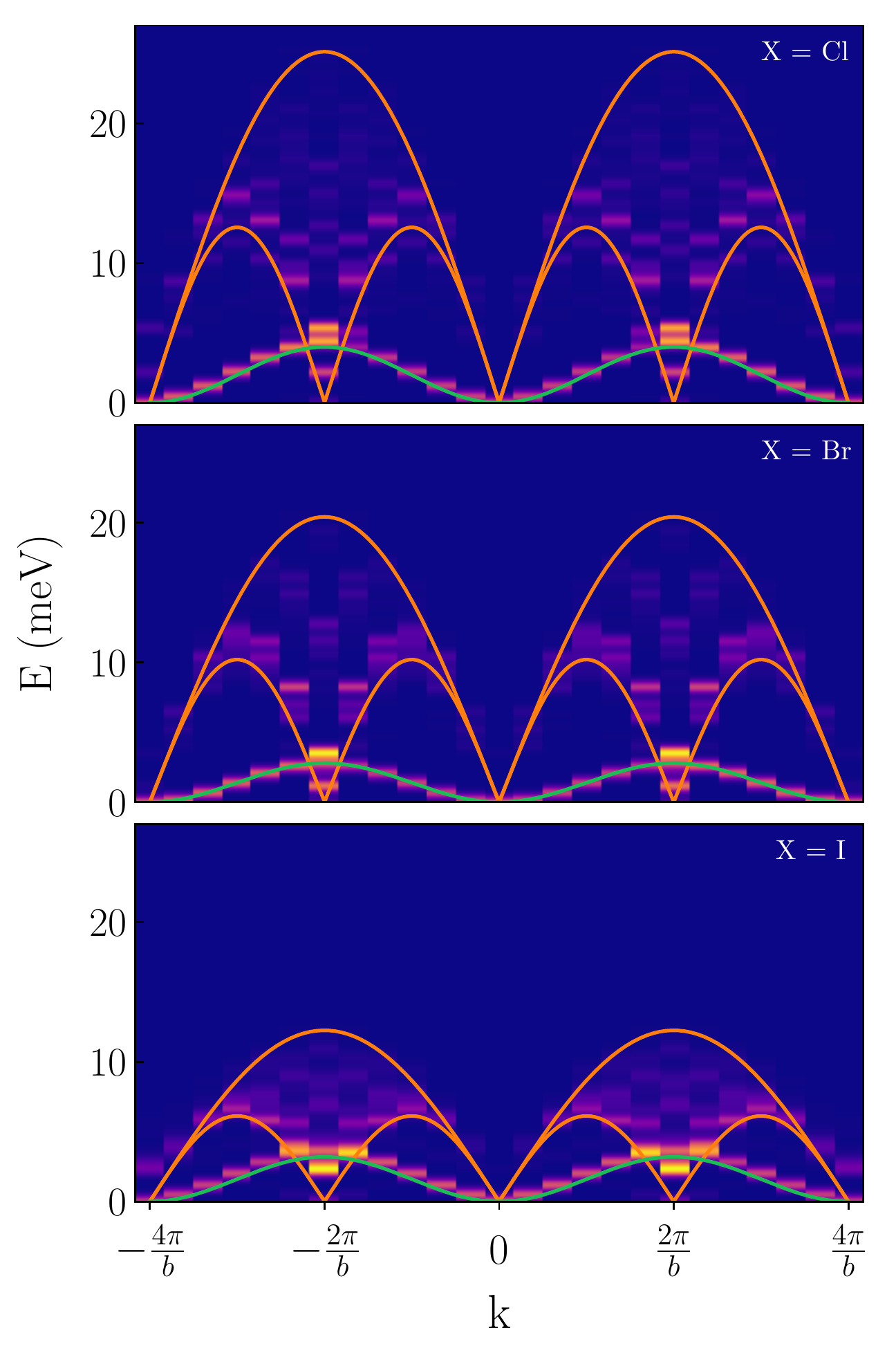}
     \caption{The dynamical structure factor $S(\mathbf{k},\omega)$ calculated using exact diagonalization for the $\mathrm{Cu_2(OH)_3X}$ compounds, where X = Cl, Br, and I. The exchanges are obtained from the experimental structures with the hydrogen atoms selectively relaxed. Also plotted are the bounds of the 2-spinon continuum (orange) obtained from the Bethe ansatz, with $J_{AFM}$ = $J_2$, as well as the magnon band (green) for a 1D ferromagnetic Heisenberg chain, with $J_{FM}$ = $J_1$. }
    \label{fig:dsf_ed_others}
\end{figure}

In addition to the above results, we repeat our calculations of $S(\vb{k},\omega)$ in the $\mathrm{Cu_2(OH)_3X}$ compounds, where X = Cl, Br, and I. In each compound, we have selectively relaxed the hydrogen atoms before obtaining the exchange constants in DFT \cite{Note2}. The results for the dynamical structure factor in each compound are shown in Fig. \ref{fig:dsf_ed_others}, and are qualitatively similar to the plots discussed above. As before, we observe a sharp ferromagnetic band and a spinon continuum in each compound. This is especially striking in the case of the Cl compound (where $J_3$ is on the order of $J_1$), and the Br compound (where both $J_3$ and $J_4$ are on the order of $J_1$.) This continued existence of the spinon continuum in these compounds suggest that magnons and spinons may coexist for relatively large values of the interchain couplings $J_3$ and $J_4$.

In the case of $\mathrm{X}=\mathrm{I}$, $J_1$ and $J_2$ are closer in magnitude than for the other oxyhalides, resulting in a larger range of momenta for which the ferromagnetic band intersects the spinon continuum. This suggests that magnon-spinon interactions may play a larger role in this compound, perhaps leading to the increase in spectral weight near the ferromagnetic band where it crosses the continuum.

\section{\label{sec:discussion}Summary and Conclusions}

In summary, to investigate the magnetic interactions in $\mathrm{Cu_2 (OH)_3Br}$, we performed first-principles DFT calculations and found the exchange constants. Using these first-principles values, we determined the classical magnetic ground state. We then obtained the magnon dispersion and dynamical structure factor using linear spin-wave theory, which are in good agreement with experiment. We find that the existence of interchain coupling is necessary to understand the magnetic order in the system. Furthermore, we find that the interchain coupling strength can be significant while continuing to preserve flat magnon bands in the interchain direction. 

Additionally, we investigated the hypothesis that the magnetic exchange is halogen-mediated, as proposed by previous experimental studies. Through analysis of the projected density of states, spin-density plots, and calculations of various exchange constants in halogen-substituted structures, we find that the exchange is instead predominantly direct or oxygen-mediated.

We also calculated the dynamical spin structure factor using exact diagonalization to examine the effect of interchain coupling on the spinon continuum. We found that the spinon continuum continues to exist, even in the presence of nonzero interchain couplings, consistent with the continuum observed experimentally. We posit that this is due to the different energy scales of spinons and ferromagnetic magnons, as well as the relative signs of interchain couplings $J_3$ and $J_4$. This is substantiated by linear spin wave theory results, which show that the interchain bandwidths in the $a$-direction scale with $J_3-J_4$ \cite{Supplement} as long as $J_3$ and $J_4$ are not large enough to modify the classical ground state.


To the best of our knowledge, the cuprate oxyhalide botallackites are the first materials in which sharp ferromagnetic magnons coexist with deconfined spinons \cite{Zhang2020}. This botallackite structure could therefore be used as a model system to study magnon-spinon interactions. For example, it would be interesting to investigate how the spinon continuum and ferromagnetic band evolve with increasing interchain coupling, which can be experimentally induced with uniaxial pressure. Our above exact diagonalization and linear spin wave theory results suggest that the spinon continuum would continue to survive with increasing uniaxial pressure along the $a$ axis. However, as the interchain couplings become on the order of $J_1$ and $J_2$, competition between the intrachain and interchain couplings will begin to play an important role. In this regime, both strong spinon-magnon interactions and significant frustration will modify the spectrum, the lifetime of both quasiparticles, and the dynamical structure factor. While these questions are outside the scope of this letter, we hope our results provide insight for future investigations.

\begin{acknowledgments}
We acknowledge useful discussions with Xianglin Ke. This work was funded by the Department of Energy through the University of Minnesota Center for Quantum Materials, under DE- SC-0016371. We acknowledge the Minnesota Supercomputing Institute at the University of Minnesota for providing resources that contributed to the results within this paper.  
\end{acknowledgments}

\bibliographystyle{unsrt}

\end{document}